%% file: main.tex
\documentclass[sigconf]{acmart}
\acmConference[MSR 2024]{21st International Conference on Mining Software Repositories}{April 2024}{Lisbon, Portugal}
\AtBeginDocument{%
  \providecommand\BibTeX{{%
    \normalfont B\kern-0.5em{\scshape i\kern-0.25em b}\kern-0.8em\TeX}}}

\setcopyright{acmcopyright}
\copyrightyear{2023}
\acmYear{2023}
\acmDOI{XXXXXXX.XXXXXXX}

\acmBooktitle{MSR 2024: 21st International Conference on Mining Software Repositories,
 April 15--16, 2024, Lisbon, ES} 
\acmPrice{15.00}
\acmISBN{978-1-4503-XXXX-X/18/06}

\usepackage{packages}
\usepackage{utils}

\newcites{SLR}{Systematic Literature Review}

\begin{document}

\title[On the Lifecycle of ML-specific Code Smells in ML-enabled Systems]{When Code Smells Meet ML: On the Lifecycle of ML-specific Code Smells in ML-enabled Systems}


\author{Gilberto Recupito}
\orcid{0000-0001-8088-1001}
\affiliation{
  \institution{Sesa Lab - University of Salerno}
  \streetaddress{P.O. Box 1212}
  \city{Salerno}
  \country{Italy}
  \postcode{84121}
}
\email{grecupito@unisa.it}

\author{Giammaria Giordano}
\orcid{0000-0003-2567-440X}
\affiliation{
  \institution{Sesa Lab - University of Salerno}
  \streetaddress{P.O. Box 1212}
  \city{Salerno}
  \country{Italy}
  \postcode{84121}
}
\email{giagiordano@unisa.it}

\author{Filomena Ferrucci}
\orcid{0000-0002-0975-8972}
\affiliation{
  \institution{Sesa Lab - University of Salerno}
  \streetaddress{P.O. Box 1212}
  \city{Salerno}
  \country{Italy}
  \postcode{84121}
}
\email{fferrucci@unisa.it}

\author{Dario Di Nucci}
\orcid{0000-0002-3861-1902}
\affiliation{
  \institution{Sesa Lab - University of Salerno}
  \streetaddress{P.O. Box 1212}
  \city{Salerno}
  \country{Italy}
  \postcode{84121}
}
\email{ddinucci@unisa.it}

\author{Fabio Palomba}
\orcid{0000-0001-9337-5116}
\affiliation{
  \institution{Sesa Lab - University of Salerno}
  \streetaddress{P.O. Box 1212}
  \city{Salerno}
  \country{Italy}
  \postcode{84121}
}
\email{fpalomba@unisa.it}

\renewcommand{\shortauthors}{Recupito et al.}

\begin{abstract}
\textbf{Context.} The adoption of Machine Learning (ML)--enabled systems is steadily increasing.
Nevertheless, there is a shortage of ML-specific quality assurance approaches, possibly because of the limited knowledge of how quality-related concerns emerge and evolve in ML-enabled systems.
\noindent\textbf{Objective.} We aim to investigate the emergence and evolution of specific types of quality-related concerns known as ML-specific code smells, \ie sub-optimal implementation solutions applied on ML pipelines that may significantly decrease both quality and maintainability of ML-enabled systems. More specifically, we present a plan to study ML-specific code smells by empirically analyzing (i) their prevalence in real ML-enabled systems, (ii) how they are introduced and removed, and (iii) their survivability. 
\noindent \textbf{Method.} We will conduct an exploratory study, mining a large dataset of ML-enabled systems and analyzing over 400k commits about \numProjectsSampling projects. We will track and inspect the introduction and evolution of ML smells through \CodeSmile, a novel ML smell detector that we will build to enable our investigation and to detect ML-specific code smells. 

\end{abstract}




\maketitle

\section{Introduction}
\input{Sections/Introduction}

\section{Background and Related Work}
\input{Sections/RelatedWork}
\label{sec:rw}

\section{Research Method}
\label{sec:method}
\input{Sections/RMV2}

\section{Threats to Validity}
\label{sec:ttv}
\input{Sections/ThreatsToValidity}
\section{Conclusion}
\label{sec:conclusion}
\input{Sections/Conclusion}

\section*{ACKNOWLEDGMENT}
\input{Sections/ack}

\balance
\bibliographystyle{ACM-Reference-Format}
\bibliography{main_bib}

\end{document}

%% file: Sections/Introduction.tex
Machine Learning (ML) evolved through the emergence of complex software integrating ML modules, defined as ML-enabled systems~\cite{martinez2022software}. 
Self-driving cars, voice assistance instruments, or conversational agents like ChatGPT\footnote{\url{https://chat.openai.com/}} are just some examples of the successful integration of ML within software engineering projects.

However, the strict time-to-market and change requests pressure practitioners to roll out immature software to keep pace with competitors, leading to the possible emergence of technical debt~\cite{cunningham1992wycash} \ie a technical trade-off that can give benefits in a short period, but that can compromise the software health in the long run.
\textit{Code smells} is a manifestation of technical debt. They are \textit{symptoms} of poor design and implementation choices that, if left unaddressed, can deteriorate the overall quality of the system~\cite{fowler1997refactoring}. 

Sculley \etal~\cite{sculley2015hidden} showed that ML-enabled systems are incredibly prone to technical debt and code smells, raising the need for a quality assurance process for ML components.
Cardozo \etal~\cite{cardozo2023prevalence} and Van Oort \etal~\cite{van2021prevalence} argued that while the issues in those systems are emerging, there is a lack of quality assurance tools and practices that ML developers can use. 
This lack of quality management assets  stimulates the proliferation of code smells in ML-enabled systems~\cite{lenarduzziSoftware}.
Consequently, given the complex nature of those systems, new types of code smells have emerged.
Considering the aspects that ML developers face when dealing with ML pipelines, Zhang \etal~\cite{10.1145/3522664.3528620} defined a new form of code smells, \textit{AI-specific code smells} (ML-CSs).
Similarly to traditional code smells, an ML-CS is defined as a \textit{sub-optimal implementation solution for ML pipelines that may significantly decrease the quality of ML-enabled systems}.
A key example of those quality issues is using a loop operation instead of exploiting the corresponding Pandas function for data handling, leading to \textit{Unnecessary Iteration} smell~\cite{10.1145/3522664.3528620}.

While some work underlines the need to explore AI\/ML-CSs~\cite{10.1007/978-3-031-49269-3_1,10.1145/3522664.3528620}, there is still a lack of knowledge on this type of  quality issues. Among the various possible causes, we outline a lack of knowledge on how ML-CSs emerge and evolve and what motivations lead developers to introduce and remove them.
This poor knowledge significantly threatens the release of ML-CSs detectors aimed at improving the system's quality. Researchers and practitioners cannot define crucial aspects of smell detection and refactoring, such as (i) \revised{the conditions where ML-CSs are more prone to be introduced and removed}, (ii) in which stage of the development lifecycle practitioners should pay attention \eg when they introduce new features or when they fix defects, (iii) what are the practices that developers use to remove code smells, (iv) in which stage a quality assurance monitoring tool should focus on tracking the evolution of ML-CSs \eg data preparation or model training.

In this registered report, we aim to bridge this knowledge gap by describing our plan to understand the evolution of ML-CSs in ML-enabled systems.
Specifically, we will perform a large-scale \revised{mixed \textit{confirmatory} and \textit{exploratory}} study considering \numProjectsSampling projects coming from the NICHE dataset \cite{10174042} and will mine over 400k commits to analyze (i) the prevalence of ML-CSs in ML-enabled systems, (ii) when and why ML-CSs are introduced and removed, and (iii) how ML-CSs survive over time.
All the collected data, analysis scripts, and additional material will be publicly available online.

On the one hand, we believe that an improved understanding of these evolutionary aspects may provide \emph{researchers and tool vendors} with insights that might be useful to characterize the peculiarities of ML-CSs and the way practitioners currently deal with them, possibly leading to the definition of novel quality assurance mechanisms that better fit the typical lifecycle of ML-CSs, thus better-assisting developers daily.
On the other hand, \emph{practitioners} with findings that may be used to improve the quality of ML-enabled systems by removing ML-CSs through the mechanisms employed by other practitioners and that will be described as part of our work.  

%% file: Sections/RelatedWork.tex
This section provides an overview of ML-CSs and summarizes the state-of-the-art concerning how code smells have been investigated in traditional and ML-enabled systems.

\lstset{
  language=Python,
    basicstyle=\footnotesize\ttfamily,
    commentstyle=\color{codegreen},
    keywordstyle=\color{codepurple},
    numberstyle=\tiny\color{codegray},
    stringstyle=\color{blue},
    backgroundcolor=\color{white},
    breaklines=true,
    breakatwhitespace=true,
    captionpos=b,
    keepspaces=true,
    numbers=left,
    numbersep=5pt,
    showspaces=false,
    showstringspaces=false,
    showtabs=false,
    tabsize=2 
}

\begin{minipage}[c]{0.96\columnwidth}
\begin{lstlisting}[language=Python, caption=Example of Gradients Not Cleared before Backward Propagation in the Transformer project., label=Gradients,escapechar=!] 
 for step, inputs in enumerate(tqdm(eval_dataloader, desc="Iteration", disable=args.local_rank not in [-1, 0])):
    for k, v in inputs.items():
        !\colorbox{mygreen}{optimizer.zero\_grad()}!
        inputs[k] = v.to(args.device)
    outputs = model(**inputs, head_mask=head_mask)
    loss, logits, all_attentions = (
            outputs[0],
            outputs[1],
            outputs[-1],
        )
    loss.backward()  # Back propagate to populate the gradients in the head mask
\end{lstlisting}
\end{minipage}

\subsection{Background}
\label{bg}
Zhang \etal~\cite{10.1145/3522664.3528620} recently released a catalog of 22 ML-CSs by empirically analyzing white and grey literature. 
\revised{In our appendix, information on the list of ML-CSs identified by the authors, their description, the pipeline stage they affect, and the quality aspects they impact are presented~\cite{Recupito2024Appendix}. }

To provide a tangible example of ML-CS, let consider \emph{Gradients Not Cleared before Backward Propagation}. 
It refers to when a developer builds a neural network in a loop operation and does not use the function \texttt{optimizer.zero\_grad()}
to clear the old gradients at the end of each iteration.
Without this operation, the gradients will gather from all the preceding backward calls.
This situation can lead to a gradient explosion, causing a failure in the training process~\cite{wang2023GradientExplosion}.
To mitigate this smell, the function \texttt{optimizer.zero\_grad()} should be used before the backpropagation step. Listing \ref{Gradients} shows an example of \textit{Gradients Not Cleared before Backward Propagation} smell for the project \textsc{Transformers}.\footnote{\url{https://github.com/huggingface/transformers/blob/main/examples/research_projects/bertology/run_bertology.py}}
We added an extra line (in green)  to indicate how to refactor the smell as denoted in the taxonomy of Zhang \etal~\cite{10.1145/3522664.3528620}.

\subsection{Related Work}
Several studies have been carried out in the context of code smells in traditional systems~\cite{khomh2012exploratory,palomba2018diffuseness,palomba2014they,taibi2017developers} also investigating their impact during the software evolution~\cite{9825773,walter2016relationship}. 
Tufano \etal~\cite{7817894} conducted a large empirical study on when and why code smells are introduced in traditional systems, their survivability, and how developers remove them.
They discovered that code smells are mainly introduced when files are created, and only a negligible percentage of them are removed through refactoring operations. 
Their impactful contribution allowed for improving the management of traditional code smells through the implementation of automatic detection and refactoring tools.
Our work, inspired by the contribution of Tufano \etal, aims to explore the nature of ML-CSs and to improve the quality assurance process for ML-enabled systems.

In the remaining part of this section, we focus on state-of-the-art traditional code smells in ML-enabled systems because, to our knowledge, no studies explicitly focus on ML-CSs in the context of ML-enabled systems.
Tang \etal~\cite{9401990} conducted an empirical study analyzing 26 machine learning (ML) projects. They identified several code anti-patterns ML-Specific, highlighting the high prevalence of \textit{Duplicated Code}. Their findings shed light on unique challenges and debt types specific to ML projects, helping researchers and practitioners understand the nature of technical debt in ML projects.
Van Oort \etal~\cite{van2021prevalence} conducted an empirical study on code smells by analyzing 74 open-source machine learning projects using PyLint. Also, in this case, they found that \textit{Duplicated Code} is the most frequent smell. Furthermore, they noticed that the emergence of code smells is more frequent in machine-learning systems than in traditional ones. 
Inspired by the work of Van Oort \etal~\cite{van2021prevalence}, Giordano \etal~\cite{giordano2021understanding} performed a large study on the diffusion of code smells over time in ML-enabled systems, focused on the activities that lead developers to introduce code smells and the survival time. The findings suggested that the smell variation does not follow a specific pattern over time; their introductions are principally due to evolutionary activities, and code smells can survive even for several years.
These previous findings were confirmed by Cardozo \etal~\cite{cardozo2023prevalence}, who, in 2023, investigated the presence of code smells by considering 29 reinforcement learning (RL) projects. Still, in this case, the results seem to go in the same direction, pointing out that the emergence of traditional code smells is more frequent in reinforcement learning projects than in traditional ones. 

Compared to previous work, our study will not focus on traditional code smells but on ML-specific code smells.
Our results could shed light on their prevalence, introduction, removal, and survivability for eliciting ways to prevent developers from introducing such smells and help remove them when already present.

%% file: Sections/RMV2.tex
The following section presents the design of the study, highlighting the main goal and the relative research questions.
We will follow the guidelines by Wohlin \etal~\cite{wohlin2012experimentation} and the ACM/SIGSOFT Empirical Standards\footnote{Available at \url{https://github.com/acmsigsoft/EmpiricalStandards}}; in particular, we will use the \quoted{General Standard}, \quoted{Data Science}, and \quoted{Repository Mining} guidelines.

\subsection{Goal and Research Questions}
The study aims to explore to what extent ML-CSs are prevalent in ML-enabled systems, when and how they are introduced and removed, and for how long they survive.
To address our goal, we formulated the specific goal through the GQM approach~\cite{caldiera1994goal}.

\goal{

\textbf{Purpose:} Explore

\textbf{Issue:} (i) the prevalence, (ii) the introduction, (iii) the removal, and (iv) the survival

\textbf{Object:} of ML-specific code smells in ML-enabled systems

\textbf{Viewpoint:} from the points of view of ML developers.
}

\begin{figure*}[ht]
    \centering
    \includegraphics[width=.70\textwidth]{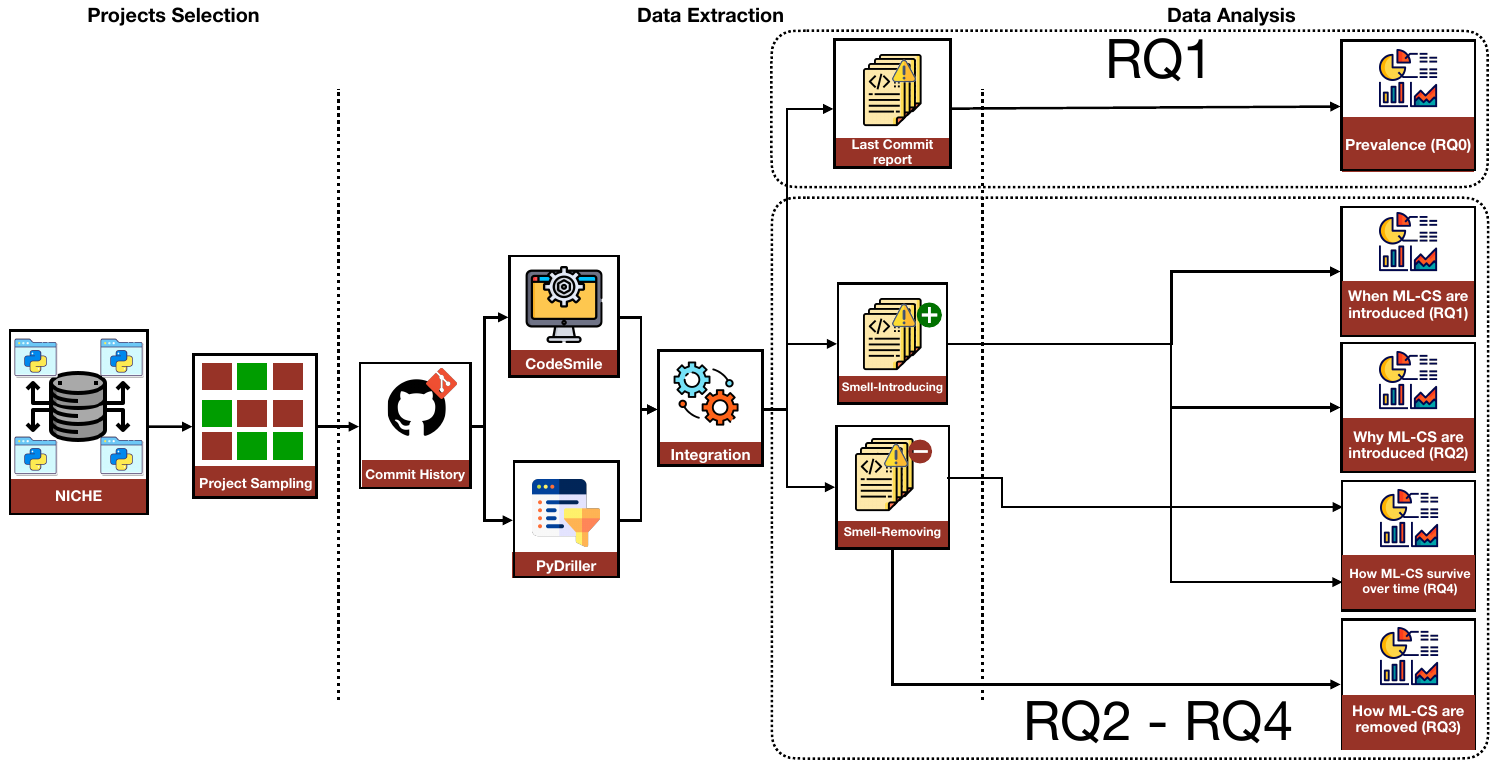}
    \caption{The process designed for the study.}
    \label{fig:process}
\end{figure*}

\Cref{fig:process} depicts the process we will follow to address our research goal by addressing a preliminary and three main research questions.

\resquestion{0}{How are ML-specific code smells prevalent in ML-enabled systems?}

The reason behind this preliminary investigation is twofold. On the one hand, we may assess the relevance of the problem: should we identify a poor prevalence of ML-CSs, this may indicate that the problem is not as relevant as in traditional systems~\cite{palomba2018diffuseness, bessghaier2020diffusion}, possibly not motivating further research on the matter. On the other hand, we may identify the most common ML-specific code smells and in which stage of an ML pipeline they manifest themselves.

\resquestion{1}{When are ML-specific code smells introduced in ML-enabled systems?}

This research question allows the classification of the conditions and contexts that lead developers introduce ML-CSs, other than the understanding if ML-CSs are injected, \eg when the ML projects are created or during the evolution of the system. The results to \textbf{RQ$_1$} would inform when ML-CSs should be mitigated.

\resquestion{2}{\revised{What tasks were performed when the ML-CSs were introduced?}}

After understanding when ML-CSs are injected, it is necessary to extract information about the reasons that led developers to update the system by introducing an ML-CS. So, \textbf{RQ$_2$} aims to extract the actions performed by developer which likely introduce ML-CSs.

\resquestion{3}{When and how ML-specific code smells are removed in ML-enabled systems?}

\textbf{RQ$_3$} focuses on the timing and methods employed to remove ML-CS.
This research question is motivated by the need to understand the strategies and circumstances under which developers address ML-CSs. 
By identifying the set of strategies that developers use to remove ML-CSs, we aim to extract insights to define automatic refactoring strategies for ML-CSs that developers would be inclined to integrate into the development processes. 

\resquestion{4}{\revised{How long do ML-specific code smells survive in the code?}}

Finally, \textbf{RQ$_4$} aims to observe the survival time of each ML-CS to identify the ones that persist in the project over time.
The outcome of the analysis can be utilized to focus on detecting the ML-CSs that exhibit greater endurance during software maintenance.

\begin{table}[t]
\centering
\footnotesize
\caption{Descriptive statistics of the NICHE projects.}
    \begin{tabular}{|l|r|r|r|}
    \rowcolor{arsenic}
    \textcolor{white}{\textbf{}} & \textcolor{white}{\textbf{Stars}} &  \textcolor{white}{\textbf{Commits}} &  \textcolor{white}{\textbf{LOC}}\\
    \hline
    Min & 100 & 100 & 10\\
    \hline
    \rowcolor{gray10} 1st Q. & 211 & 219 & 3,829 \\
    \hline
    Median & 529 & 420 & 9,235 \\
    \hline
    \rowcolor{gray10} Mean & 1,978 & 1,307 & 24,414 \\
    \hline
    3rd Q. & 1,641 & 1,070 & 21,845 \\
    \hline
    \rowcolor{gray10} Max & 76,838 & 90,927 & 699,513\\
    \hline
    \end{tabular}
\label{table:DescriptiveStatistics}
\end{table}

\subsection{Dataset Description and Projects Selection}
\label{sec:dataset}
We will rely on the NICHE dataset~\cite{10174042} for our investigations.
This dataset contains \numNicheProjects ML-enabled systems and was released at MSR '23.
We selected it for two reasons.
On the one hand, it contains only popular, active ML projects with extensive commit histories \ie projects with over 100 stars on GitHub, with commits more recent than May 1$^{st}$, 2020, and with at least 100 commits, allowing us not to select personal or inactive projects.
On the other hand, it contains heterogeneous projects with different characteristics.

To verify the feasibility of the analysis on the selected dataset, we preliminary mined it.
This operation was necessary because it is reasonable to suppose that some projects could be no longer available for some reason (\eg repositories are archived, some communities migrated to other version control systems, or some repositories have restricted access).
At the end of this step, we identified \numActiveProjects projects available out of the \numNicheProjects and \numTotalCommits commits. 
\Cref{table:DescriptiveStatistics} shows the descriptive statistics on the variables \quoted{Stars}, \quoted{Commits}, and \quoted{Lines of Code} provided in NICHE~\cite{10174042}.
As we can notice from the metrics extracted, the project distribution in the NICHE dataset presents a median of about 529 stars, 420 commits, and 9,235 lines of code, suggesting that the projects have a high development activity.

Observing the statistics of the projects, we noticed a high variability between projects in terms of lines of code (LOC).
According to Zhou \etal~\cite{4967613}, the project size is an impactful confounding variable when analyzing code-related aspects.
Therefore, we analyzed the active projects in the dataset through a percentile distribution analysis and divided them into three groups:

\begin{description}[leftmargin=0.3cm]
    \item[Small:] Projects with a number of lines of code below the 30$^{th}$ percentile. This set consists of 167 projects, all containing less than 4,765 lines of code. 
    \item[Medium:] Projects with a number of lines of code above the 30$^{th}$ percentile and below the 60$^{th}$ percentile. This set consists of 169 projects, all containing less than 11,836 lines of code.
    \item[Large:] Projects with a number of lines of code above the 60$^{th}$ percentile. This set consists of 224 projects, all containing more than 11,836 lines of code.
\end{description}

\begin{table}[t]
\centering
\footnotesize
\caption{Descriptive statistics of the active projects divided by size.}
    \begin{tabular}{|l|l|r|r|r|}
    \rowcolor{arsenic}
    & & \textcolor{white}{\textbf{Stars}} & \textcolor{white}{\textbf{Commits}} & \textcolor{white}{\textbf{LOC}}\\
    \hline
        \multirow{6}{*}{Small} & Min & 100 & 100 & 1,648\\
        \hhline{~|-|-|-|-|}
        & \cellcolor{gray10}1st Q. & \cellcolor{gray10} 171 & \cellcolor{gray10} 148 & \cellcolor{gray10} 2,301 \\
        \hhline{~|-|-|-|-|}
        & Median & 367 & 244 & 2,921 \\
        \hhline{~|-|-|-|-|}
        & \cellcolor{gray10}Mean & \cellcolor{gray10} 1,170 & \cellcolor{gray10} 552 & \cellcolor{gray10} 3,115 \\
        \hhline{~|-|-|-|-|}
        & 3rd Q. & 879 & 436 & 3,871 \\
        \hhline{~|-|-|-|-|}
        & \cellcolor{gray10}Max & \cellcolor{gray10} 13,265 & \cellcolor{gray10} 13,542 & \cellcolor{gray10} 4,763\\
    \hline
        \multirow{6}{*}{Medium} & Min & 100 & 105 & 4,783\\
        \hhline{~|-|-|-|-|}
        & \cellcolor{gray10}1st Q. & \cellcolor{gray10} 159 & \cellcolor{gray10} 241 & \cellcolor{gray10} 6,268 \\
        \hhline{~|-|-|-|-|}
        & Median & 290 & 375 & 7,817 \\
        \hhline{~|-|-|-|-|}
        & \cellcolor{gray10}Mean & \cellcolor{gray10} 1,138 & \cellcolor{gray10} 610 & \cellcolor{gray10} 8,039 \\
        \hhline{~|-|-|-|-|}
        & 3rd Q. & 907 & 722 & 9,344 \\
        \hhline{~|-|-|-|-|}
        & \cellcolor{gray10}Max & \cellcolor{gray10} 18,087 & \cellcolor{gray10} 3,299 & \cellcolor{gray10} 11,835\\
    \hline
        \multirow{6}{*}{Large} & Min & 105 & 103 & 12,005\\
        \hhline{~|-|-|-|-|}
        & \cellcolor{gray10}1st Q. & \cellcolor{gray10} 336 & \cellcolor{gray10} 405 & \cellcolor{gray10} 17,656 \\
        \hhline{~|-|-|-|-|}
        & Median & 901 & 855 & 27,352 \\
        \hhline{~|-|-|-|-|}
        & \cellcolor{gray10}Mean & \cellcolor{gray10} 3,052 & \cellcolor{gray10} 2,271 & \cellcolor{gray10} 54,342 \\
        \hhline{~|-|-|-|-|}
        & 3rd Q. & 2,652 & 1,514 & 46,488 \\
        \hhline{~|-|-|-|-|}
        & \cellcolor{gray10}Max & \cellcolor{gray10} 33,741 & \cellcolor{gray10} 47,094 & \cellcolor{gray10} 661,808\\
    \hline
    \end{tabular}
\label{table:DescriptiveStatisticsActiveProjects}
\end{table}

Due to the potential computational issues arising from the large number of projects and commits, we will apply a statistically significant sampling for each group. Specifically, we will select the projects considering for each population, a sample with 95\% confidence level and 5\% margin of error.
As a result, we will consider 117 projects, composed of 64,607 commits, for Small projects, 118 projects, composed of 71,380 commits, for Medium projects, and 142 projects, composed of 265,671 commits, for the Large projects \ie we want to analyze \numProjectsSampling projects and \numCommitsSampling commits.
\Cref{table:DescriptiveStatisticsActiveProjects} shows the descriptive statistics for each size group. 

\revised{In addition to analyzing projects based on their size in terms of LOC, we will also consider the most related characteristics that led the authors to define a project as ML-engineered.
One such characteristic is the adoption of Continuous Integration (CI).
The presence of a CI pipeline may directly affect the quality assurance mechanisms implemented by the projects, possibly affecting the presence of ML-CSs. This distinction between projects with and without CI allows us to explore potential differences in the prevalence and management of ML-CSs between the two groups. Therefore, to incorporate this aspect into our analysis, we thoroughly examined the dataset and found that 319 projects utilize CI tools, whereas 247 projects do not incorporate CI into their development environment.}

\subsection{Data Extraction}
After cloning the projects, we will gather fine-grained structural metrics using PyDriller, a framework helpful to analyze Git repositories~\cite{spadini2018pydriller}. We will extract the commit history of a project \textsc{P} belonging to the selected projects. For each commit, C$_i$ $\in$ P, we will collect the total number of files, the number of removed and added files, the commit date, and the commit message.

\subsection{ML-Specific Code Smell Detection}
To achieve the study's objectives, we will develop an ML-CS detection tool, \CodeSmile.
This tool will analyze the Abstract Syntax Tree (AST) of code to gather information about statements and methods.
Notably, despite the clarity and specificity of ML-CS definitions provided by Zhang \etal~\cite{10.1145/3522664.3528620}, there is a lack of automated solutions for detecting these code smells.
\revised{\CodeSmile aims to fill this gap and provide an automatic solution to detect a set of 14 ML-CS; for the sake of space limitations, the definitions of the targeted ML-CS are reported as part of our online appendix~\cite{Recupito2024Appendix}. 
The tool will use the rule-based conditions defined by Zhang~\etal~\cite{10.1145/3522664.3528620} to identify ML-CSs effectively, and we will manually validate it by selecting a statistically significant sampling of smells. The feasibility of static analysis to detect these smells is discussed in the technical report available in our online appendix~\cite{Recupito2024Appendix}.}
The tool will use the rule-based conditions defined by Zhang~\etal~\cite{10.1145/3522664.3528620} to identify ML-CSs effectively, and we will manually validate it by selecting a statistically significant sampling of smells.

\revised{The validation set will be defined as follows. Starting from the whole set of files included in the software projects considered in the study, we will first identify the files containing ML modules - in this way, we will filter out those files that, by definition, cannot contain any ML-CS. This first step will be performed by statically analyzing the content of each file: if it contains a reference to libraries such as Pandas,\footnote{\url{https://pandas.pydata.org/}} TensorFlow,\footnote{\url{https://www.tensorflow.org/?hl=it}} Theano,\footnote{\url{https://pypi.org/project/Theano/}} or PyTorch,\footnote{\url{https://pytorch.org/}} then the file will be marked as ML-related. The choice of the libraries to verify is mainly based on previous work \cite{10174042}: the authors of the NICHE dataset have indeed defined ML projects as those relying on TensorFlow, Theano, or PyTorch. We also consider Pandas because several of the code smells targeted by our study refer to the suboptimal use of this library. We will then pick a statistically significant sample (confidence level = 95\%, margin of error = 5\%), which will finally form our validation set.}

\revised{Upon completing the step above, we will then recruit external ML engineers, asking them to inspect the files in the validation sample (or a part thereof, in case the number of files to validate would be excessively large) and annotate the ML-CS instances they contain. In particular, for each ML engineer, we will prepare a validation package containing (1) the set of files to analyze, (2) a readme file reporting definitions and examples of the specific code smells we are interested in, and (3) a spreadsheet containing a number of rows equals to the files to validate and a number of columns equals to the smells to evaluate. Given this validation package, the ML engineers will be asked to fill each entry of the spreadsheet with a \quoted{Yes} if the \textit{i-th} file is affected by the \textit{j-th} code smell, with a \quoted{No} otherwise. We will give ML engineers up to 21 days to send us back the annotated spreadsheet. The number of ML engineers involved will depend on the validation sample size. We will start recruiting ML engineers from our contact network. Whenever needed, we will attempt to involve further practitioners through public calls on social media, practitioner’s blogs, and specialized ML platforms. Among all the candidates, we will retain only those having at least three years of experience in the development of ML systems.}

\revised{Once we receive the annotated spreadsheets, we will analyze them as follows. On the one hand, we will compute Cohen’s k inter-rater agreement to measure the level of agreement among the ML engineers that assessed the smelliness of the same files - this will be useful to understand the extent to which ML code smells are perceived as such. On the other hand, we will define a ground truth by means of majority voting: for each file of the validation set, we will finally consider it as affected by a given smell if the majority of the ML engineers marked it as smelly. Such a ground truth will finally be used to assess the capabilities of CodeSmile in terms of precision and recall, hence assessing the amount of false positive and negative output by the tool.}

As a result of the detection, the tool will report all the identified ML-CS  with the relative position.
\CodeSmile will successively analyze each commit of ML projects to report all the information helpful in analyzing introduction, survival time, and removal of ML-CSs.
To conduct our analyses, we will combine this data with those previously extracted using PyDriller~\cite{spadini2018pydriller}.

\subsection{Commit Data Extraction}
\label{sec:commit_data}

After identifying the method for detecting ML-CS, we will extract the commit data for each project to address our research questions.
First, we will identify the smell-introducing and smell-removing commits for each identified AI-CS instance.
Specifically, we will track each smell $s_i$ identified in a commit $c_i$, using its file name and line number.
We will analyze the project's history from the first commit, comparing $c_i$ and $c_i+1$ pairwise.
For each pair of consecutive commits, we will consider the two following cases:

\begin{enumerate}
\item If $c_i+1$ contains a smell $s_i$ not contained in $c_i$, then $c_i+1$ is the smell-introducing commit for $s_i$.
\item If $c_i$ contains a smell $s_i$ not contained in $c_i+1$, then $c_i+1$ is the smell-removing commit for $s_i$.
\end{enumerate}


After collecting smell-introducing and smell-removing commits, we will analyze the commit messages to understand the rationale behind introducing and removing ML-CSs.


\subsection{Data Analysis}
The following section explains how we want to analyze the collected data to respond to our research questions.

\paragraph{\emph{\textbf{RQ}$_0$}: How are ML-specific code smells prevalent in ML-enabled systems?}
\CodeSmile will analyze \revised{last snapshot of } the selected ML-enabled systems to observe the prevalence distribution of each ML-CS.
Statistical descriptions and plots will be employed to understand the characteristics of each distribution.
Then, insights on the most prevalent ML-CS across several projects will be provided. 
Such results will be further enriched by mapping each identified ML-CS to the related ML-pipeline stage, utilizing the mapping framework established by Zhang~\etal~\cite{10.1145/3522664.3528620}.
This additional mapping step will enhance our understanding of ML-CSs, revealing the most prone areas within the ML development pipeline.
Each analysis will be conducted considering the effect that related factors could have.
Through the different size groups \revised{and the adoption of continuous integration} defined in \Cref{sec:dataset}, we will apply statistical tests to understand whether these are possible factors influencing the prevalence of ML-CS in ML-enabled systems. \revised{At first, we hypothesize that different types of ML code smells may differ in terms of prevalence. Hence, we formulated the following null hypothesis:}
\revised{
\begin{description}[leftmargin=0.3cm]
\item[H0:]\textit{ There is no statistically significant difference between the prevalence of the smells i and j.}
\end{description}}
\revised{with i and j belonging to the set of ML smells S considered in the study.}
\revised{Secondly, we hypothesized that the prevalence of ML code smells may depend on the size of the ML projects. Larger projects may indeed be more complex and involve more contributors, increasing the likelihood of introducing code smells during development. As such, we formulated the following null hypothesis:}
\revised{
\begin{description}[leftmargin=0.3cm]
\item[H1:] \textit{There is no statistically significant difference in the prevalence of the smell i among large, medium, and small projects}
\end{description}
}
\revised{with i belonging to the set of ML smells S considered in the study, large projects being those having a size (in terms of lines of code) above the 60$^{th}$ percentile of the distribution of the sizes of all projects, medium projects being those having a size between the 30$^{th}$ and 60$^{th}$ percentile of the distribution of the sizes of all projects, and small projects being those having a size lower than the 30$^{th}$ percentile of the distribution of the sizes of all projects.}

\revised{We hypothesized that projects relying on a Continuous Integration (CI) pipeline may have a lower prevalence of code smells than those not relying on that. Indeed, the presence of a CI pipeline may have a direct effect on the quality assurance mechanisms implemented by the projects, possibly affecting the presence of ML code smells. Hence, we formulated our last null hypothesis:}
\revised{
\begin{description}[leftmargin=0.3cm]
\item[H2:] \textit{There is no statistically significant difference in the prevalence of the smell i between projects relying and not on a Continuous Integration pipeline.}
\end{description}
}
\revised{with i belonging to the set of ML smells S considered in the study.}

\revised{For each null hypothesis, we also defined an alternative hypothesis. Regarding statistical verification, we plan to use different tests for the three hypotheses we formulated. For \textit{H0} and \textit{H2}, we will use the non-parametric Wilcoxon test~\cite{conover1999practical}, which investigates significant differences between two populations. Then, for the analysis for \textit{H1} and given the goal of exploring differences between three groups, we will use a test that allows us to study differences across more than two populations: the non-parametric Friedman test.
}

The results will be statistically significant at $\alpha$=0.05.
We will normalize the data distribution by the project LOC to avoid possible biases and quantify the effect size using the Cliff's Delta ($\delta$)~\cite{Cliff1993DominanceSO}.



\paragraph{\emph{\textbf{RQ}$_1$}: When are ML-Specific code smells introduced in ML-enabled systems?}
After the commit data extraction phase described in \Cref{sec:commit_data}, we will collect all the smell-introducing commits to understand when each ML-CS is introduced.
The outline of the smell-introducing commits will allow us to understand which ML-CSs are introduced during file creation and which occur during the evolution and maintenance of ML projects.
\revised{To gain insights into the lifecycle of ML-specific code smells, we will also implement a segmentation approach based on three key factors: development time, activity levels, and distance from the release. 
The first two segments will be used to examine the moment at which ML-CSs are introduced. Each commit is categorized based on its duration since the project started and its position in the commit history. Subsequently, we will conduct an analysis within each segment to determine the presence of smell-introducing commits. In addition to these segments, the third segment investigates the relationship between the introduction of ML-CSs and project releases. We identify commits labeled as "Release" using PyDriller and categorize all other commits based on their proximity to the subsequent project release (\eg one day before the next release).
By examining the temporal proximity of code smell occurrences to release events, we aim to ascertain whether the timing of releases influences developers' proneness to introduce ML-CSs. 
Table \ref{table:introduction_segmentation} provides the segments and the value that will be used for the segmentation.}
\begin{table}[ht] 
\caption{Segmentation of commits for the analysis of the introduction of ML-CSs.}
\centering
\footnotesize
    \begin{tabular}{|l|p{0.3\columnwidth}|p{0.25\columnwidth}|}
    \rowcolor{arsenic}
    \textcolor{white}{\textbf{Tag}} & \textcolor{white}{\textbf{Description}} &
    \textcolor{white}{\textbf{Values}}\\
    \hline
    Development Time & Based on the duration since the project's starting date & [one week, one month, one year, more than one year] \\
    \hline
   \rowcolor{gray10}Activity Level & Based on its sequence in the project's commit history, identifying the number of previous commits. &  [first 10\% of commits, first 20\% of commits, first 50\% of commits, after the first 50\% of commits] \\
    \hline
    Distance from a Release: & Based on the time elapsed before the next release. &  [one day, one week, one month, more than one month] \\
    \hline
    \end{tabular}
\label{table:introduction_segmentation}
\end{table}

\paragraph{\emph{\textbf{RQ}$_2$}: \revised{What tasks were performed when the ML-Specific code smells were introduced?}} 
After collecting the list of smell-introducing commits for all ML-CS instances, we will analyze their messages to explain the rationale behind the changes.
Specifically, we will leverage pattern matching, as previously done by Tufano \etal~\cite{7817894} to analyze why traditional code smells are introduced.
In detail, we will extract the rationale, starting from the label set indicating the change operations described in \Cref{table:change_operations}.
Finally, we will analyze to what extent commit rationales and introduced ML-CSs co-occur.
We will analyze whether the smell-introducing commits acknowledge the presence of the ML-CS, resulting in self-admitted ML-CSs.

\begin{table}[ht]
\caption{Change operation tags for the rationale analysis.}
\centering
\footnotesize
    \begin{tabular}{|l|p{0.7\columnwidth}|}
    \rowcolor{arsenic}
    \textcolor{white}{\textbf{Tag}} & \textcolor{white}{\textbf{Description}}\\
    \hline
    Bug Fixing & The commit aimed at fixing a bug.\\
    \hline
    \rowcolor{gray10}Enhancement & The commit aimed at implementing an enhancement in the system.\\
    \hline
    New Feature & The commit aimed at implementing a new feature in the system.\\
    \hline
    \rowcolor{gray10}Refactoring & The commit aimed at performing refactoring operations.  \\
    \hline
    \end{tabular}
\label{table:change_operations}
\end{table}

\paragraph{\emph{\textbf{RQ}$_3$:} When and how ML-specific code smells are removed in ML-enabled systems?}
After analyzing the conditions and reasons for introducing ML-CSs, we will perform a similar analysis from the smell-removing commits. 
\revised{As for \textbf{RQ}$_1$, we will first verify whether ML-CSs are mitigated in a smell-removing commit and identify which are unmitigated, employing the same segmentation adopted and represented in Table \ref{table:introduction_segmentation}.}
Afterward, we will focus on the smell-removing commits. 
We will collect the messages of all smell-removing commits using the pattern matching approach adopted in \textbf{RQ}$_2$, relying on the tags in \Cref{table:change_operations} to extract the rationale behind the removal.
This analysis will allow us to extract the refactoring operations addressing ML-CSs.
From the set of the smell-removing commits analyzed, we will consider apart the commits that do not perform changes to ML-CSs but remove them by deleting the files. 

\paragraph{\emph{\textbf{RQ}$_4$:} \revised{How long do ML-Specific code smells survive in the code?}}
To understand the lifetime of each ML-CS instance, we will compute the number of commits from the smell-introducing commit to the smell-removing commit and the time span in days.
Given such values, we will compute the mean lifespan of each ML-CS type to understand which smells survive for a longer lifespan.

\subsection{Public Data Availability}
To ensure the replicability of this work and enable researchers to build upon our study, we will release all materials, including scripts and datasets, in an online appendix hosted in permanent storage.

%% file: Sections/ThreatsToValidity.tex
This section discusses possible threats to validity that could impact our results and the strategies we will adopt to mitigate them.

\paragraph{Threats to Construct Validity}
A possible threat concerns the detection of the ML-CS instances.
Indeed, a rule-based detection tool could lead to false positives and negatives.
To limit this threat, we will implement a pattern-matching strategy using an AST and reflecting the definitions provided by Zhang \etal~\cite{10.1145/3522664.3528620}.
We will manually validate the tool's accuracy by selecting a statistically significant sampling of ML-CSs.
While this solution limits the detection of the set of smells defined in the literature, this is a starting point for creating a quality assessment tool for ML-enabled systems.

Another threat regards data collection, particularly the mismatch between the data collected and the properties of the projects.
To mitigate this threat, we will use an established tool to mine repositories \ie PyDriller, as already done in previous work ~\cite{giordano2021understanding,Riquet2022403,Rhmann202136}. 

\paragraph{Threats to Internal Validity}
In the context of survivability analysis (\textbf{RQ$_4$}), we will exclude the smells developers could not fix because of lack of time.
In other words, we will remove from our analysis smell-introducing commits too close to the last commit of the project by excluding those instances whose smell-introducing date summed to the median removal time is beyond the end of the commit history.
Another threat regards smell-removing commits.
We will consider a smell removed at commit $c_i$ when the instance is detected at commit $c_{i-1}$ but no longer detectable at commit $c_i$.
This approach could lead to some imprecision due to refactoring, not removing the ML-CS in the commit $c_i$ but modifying the source code until it no longer matches the established detection rules.

\paragraph{Threats to External Validity}
The main threat to the generalizability of the results regards the dataset we will use. 
We are conscious that the project selection is a critical experiment component.
Therefore, we will rely on the NICHE dataset~\cite{10174042},  \ie a large dataset that contains only real ML-enabled systems.
We will analyze \numProjectsSampling projects and over 400k commits, proposing a large empirical study.
The projects are provided from different contexts and have different characteristics (\eg size, number of files).
We know the results could not directly apply to industrial projects; however, we invite researchers to replicate our study on closed-source projects to identify differences and common points.

Another generalizability threat is related to the programming language used to write the systems under analysis \ie Python. We are aware that due to the specific characteristics of this programming language, the generalizability of our results needs to be confirmed by future studies using other programming languages. We intend to conduct similar investigations for other programming languages as part of our future agenda to confirm the findings. 

\paragraph{Threats to Conclusion Validity}
The main threat to validity is related to the statistical test that we want to apply to address the $RQ_0$ \ie the \revised{Friedman or the Wilcoxon test and Cliff’s Delta}, since the characteristics of the distribution of the data can violate the assumptions that need to be presented to conduct the tests.
Before applying these tests, we will verify the distribution of the projects to verify the normality of the data, and only after this will we select the most appropriate test.

As the last conclusion validity threat, to calculate the lifespan in \textbf{$RQ_4$}, we want to use the number of commits and days as a time indicator. These two indicators could not be precise enough. Due to their internal policies, some communities may not commit even over long periods, thus making comparative analyses inaccurate. 

%% file: Sections/Conclusion.tex
This paper describes a plan to investigate AI-specific code smells.
Through the analysis of 337 projects, we want to understand (i) the prevalence of AI-CSs, (ii) when and why they are introduced and removed, and (iii) their survival time.
The implications of this study could be significant for the AI engineering community.
On the one hand, we will provide \CodeSmile a tool to detect AI-CS instances.
On the other hand, the valuable insights from the large-scale empirical study we will perform will allow for eliciting ways to prevent developers from introducing such smells and help remove them when already present.
Using the findings of this study, we aim to enhance the state of knowledge in the domain of AI quality assurance, directing future research endeavors towards the identification and resolution of quality issues specific for AI-enabled systems, moving towards the improving of the overall AI quality.

%% file: Sections/ack.tex
This work has been partially supported by the European Union - NextGenerationEU through the Italian Ministry of University and Research, Projects PRIN 2022 "QualAI: Continuous Quality Improvement of AI-based Systems", grant n. 2022B3BP5S , CUP: H53D23003510006.